# Why dimensionless units should not be used in physics

## Petr Křen

Czech Metrology Institute, Okružní 31, 63800 Brno, Czech Republic
E-mail: pkren@cmi.cz

**Abstract**

The quantities of dimension one – known as the dimensionless quantities – are widely used in physics. However, the debate about some dimensionless units is still open. The paper brings new interrelated arguments that lead to the conclusion to avoid physical dimensionless units, except one for the mathematical multiplication identity element that should not be introduced into a system of physical units. It brings the coherence to the International System of Units (SI) and it will remove ambiguities rising from the conflict between the mathematical properties and the physical conventions.

**Keywords:** dimensionless, unit, quantity, mole, radian

## 1. Introduction

Any physical quantity must be expressed as a number and a reference. The quantity and the reference have a physical meaning, while the numerical value has a mathematical origin. The metrological notation for a quantity $q$ gives an equation

$$q = \{q\}[q], \qquad (1)$$

where $\{q\}$ is the numerical value and $[q]$ is the unit. Eq. (1) can be rewritten in terms of the measurement output, where the numerical value can be defined as

$$\{q\} = \frac{q}{[q]}. \qquad (2)$$

Now the mathematical part is on the left-hand side of eq. (2), whereas the physical part is on the right-hand side. The numbers are mathematical objects independent of physical reality. The numbers are the results of measurements that can be processed by means of mathematics that is separated from the physical realisations of units. The physical quantities are also not involved in an axiomatic system of mathematics. This also corresponds to the convention for axis labels e.g. in the *Metrologia* journal that are written in the form $q/[q]$, and thus there are only numbers plotted in mathematical graphs with the labels specifying their physical meaning. Note that the fixings of numerical values in the new SI definitions of the base units lead to the fact that these units are defined as constant quantities [1]. Historically, it was for example a ratio $\{L\}$ of the measured length $L$ to the length of International Prototype Metre $\mathfrak{M}$, written as $L(\mathfrak{M}) = $ m (that is, the metre is a quantity of the "constant" Metre), and thus defined as the unit $[L]$. The traceability chain was realized as multiplications of numerical factors obtained from calibrations to calculate the final factor (the numerical value) of measured $L$ to the physical unit represented by $\mathfrak{M}$ as an indirect comparison. By analogy, the International Prototype of the Kilogram $\mathfrak{K}$ had mass $m(\mathfrak{K}) = $ kg and was assumed as a constant quantity. Currently, the fixings of numerical values of some physical constants that represent definitions of the new SI units allow direct traceability between quantities by constant factors to the unit of time $9192631770 \cdot t(\text{Cs}) = \text{s} = [t]$. Thus, we must keep the mathematical numerical values separately from physical values of quantities and units. That is, we must use the system where all mathematical factors are dimensionless and all physical units have a physical dimension. And we will see below, why this formal requirement is so useful and correct. Note that historically the SI system of units was preceded by a set of units. Thus, some of these units were also inconsistent. Thus, some of these units were also inconsistent and did not form a system. For example, the separate realization-based definitions of the volt, the ohm and the ampere were incompatible. These three units were defined by different practical realizations and numerical factors, and thus they did not match Ohm's law exactly. However, the SI system of units evolved and now it is closer to a state without such inconsistencies. However, some steps are still needed and they can be done together with the expected change in the definition of the SI second.

The paper goes through a set of old and new arguments that must be mentioned explicitly to demonstrate that the proposal about the dimensionless units does not have a loophole. In section 2, the integer numbers and related units are discussed, especially the mole. In section 3, an analogical treatment is used for the real numbers and the angle. In section 4, the ambiguity of identity element is examined.

## 2. Numbers

The integer numbers are enumeratively defined in mathematics. Their set is infinite and they have their own mathematical symbols. For example, the diameter $d$ is expressed as 2-fold of the radius $r$ by the following equation

$$d = 2r, \qquad (3)$$

where 2 represents the number two and the multiplication sign is conventionally omitted. If the factor in unknown, than a new symbol (e.g. $N$) is used. A number that is substituted by the symbol $N$ can vary, and thus $N$ is named the variable. Note that not all symbols need to be a physical quantity. The $N$ is still a (variable) number and it is not a physical quantity. Then we can write

$$d = Nr. \qquad (4)$$

The corresponding unit equation for the SI units is

$$\mathrm{m} = 1 \cdot \mathrm{m} = \mathrm{m}, \qquad (5)$$

where any numerical factor is conventionally omitted. The quantities diameter and radius are generally considered to be so-called quantities of the same kind [2], and thus they are also additive. For example, we can add radiuses and diameters of different circles to express some total distance.

The old SI definition of the mole was already subject of doubts. Nevertheless, the old definition of the mole was related to the mass. However, the new SI definition of the mole is directly based on a number only that strengthen these doubts [3-4]. The "numbers of entities are quantities of dimension one" and also according to the International Vocabulary of Metrology (VIM) [2]. That is, they are dimensionless for general entities. However, the mole, which is the base unit of amount of substance, seems to be a number of entities in substance. The old SI definition had the form that relied on the unit of mass represented by the prototype and on the material property of the specific isotope of carbon. In the new SI definition, however, it is the fixed number (numerical value of the Avogadro constant $N_A$), whose use is restricted for specified entities (e.g. particles) [5]. One can ask, "How is it possible that the number related to the general entity is dimensionless, while the same number can have a physical dimension for a specific entity?" The answer might be that it is due to some historical reasons. It originates in the large number that connects the macroscopic (old Pt-Ir kilogram) and the microscopic world (particles) and it cannot be counted by any current technological means in physics. However, it is still a number (according to the VIM) and it is not a new dimension (according to the new SI system of units). Note that the amount of substance has its own dimension in dimensional analysis of quantities. However, the mole is based only on the dimensionless number. Although the Dirac large number hypothesis [6] inspired some people to the physical compactification associated in a dimensional change, it has not yet been proven and the gravitational constant $G$ is not a part the SI base unit definitions [7]. It sounds strange that we can select an integer number, and then by definition it becomes a physical unit with a dimension. The physical meaning begins e.g. with an introduction of a mass constant $m_e$ (of a physical unit with dimensions) for a given kind of entity (particles) that is multiplied by this numerical factor symbolized by mol, and thus the result is the mass $M$ (the quantity of the same kind with same units) as follows

$$M = N_A m_e. \tag{6}$$

Nevertheless, $N_A$ represents integer number

$$M = 602\,214\,076\,000\,000\,000\,000\,000\, m_e \tag{7}$$

that is clearly mathematical as well as 2 in eq. (3) or any other number. The conventional restriction to use a numerical value only for selected (a subset of) entities offers a non-systematic interpretation with a physical meaning. However, the restriction is not necessary from mathematical reasons.

The dozen that corresponds to the value equal to 12 is a numerical factor and it is not standardly viewed as a unit. It can be used for a construction of the duodecimal numeral system using the gross that corresponds to the dozen of dozens. The number of an entity can be e.g. 5 dozens as well as 5 moles which is conventionally restricted only to physical particles. Thus, we can have five dozens of apples or five moles of particles. However, the entity must be specified in the definition of a quantity, and then, it is included in the symbol of quantity and it is not present as a unit (or its symbol). For example, the number of apples $N_{apple}$ is 5 dozens and it can be written as

$$N_{apples} = 60 \tag{8}$$

in our case. That means, it is without any unit like "apple" on the right-hand side of eq. (8). The number of oranges $N_{oranges}$ must also be defined in a similar way and if we want to count them together, then we must define some quantity like $N_{fruits}$. However, we will not define new units. For example point, line, graduation, cycle, bit, trit, qubit, gate, event, money, particle, pixel (and their derived "units" - e.g. Mcyc/s, Gb/s, events/year, births/month, USD/week, particles/s, particles/m$^3$, pixels/m$^2$) are not units. They are only entities that can be counted (within a specified time interval or a specified space) and any entity must be omitted when we write units (e.g. 1/s, 1/m$^3$, etc.). None of these entities are the SI units. And thus, by analogy, the mole is not a unit and the restriction to the particles must be generally included only in the definition of quantity and its symbol. Therefore, the Avogadro constant can be used (using some symbol) as a dimensionless numerical factor for practical reasons. However, the mole as a unit should not be used and it is not necessary that it exists having the status of a unit. The change of its status cannot affect its level of practicality in usage, and thus the practicality cannot be an argument against such a proposal. A historically accepted dimensional confusion in naming illustrates also e.g. mile (originally meaning only 1000) as the simplified name of unit for the length. That is, there was not a sufficient pressure for a proper differentiation of dimensionless units in the past. And we can see potential problems e.g. the proposal that the kilomole should be the SI base unit in order to avoid factors such as 1000 [8].

As an additional argument it must be noted that the new SI system has one important property. All physical measurements can be reduced to a counting of specified events in a time interval, and then the measured values will be calculated. However, the mole is the only one base unit of the new SI system that does not depend on the unit second within the definitions. The mole is also the only one base unit that has not a definition-based relation to another base unit. Consequently, the mole forms an independent base that constitutes an island (see Fig. 1). Of course, the definition of the mole is based on the selected number (that cannot be directly counted for its large value) and it has not a relation to any physical phenomenon (it is out of the time) contrary to the rest of the base units. And thus it is staying outside a united physical system of units ("in its own extra dimension") as well as any other number (like the dozen) that cannot be accepted as a physical unit. The existence of the mole as the base unit also does not allow a full conversion of the SI system to any natural system of physical units because the number in the new SI definition of the mole cannot be equalized to 1 because the natural mole will be the "uno-like" unit that will be discussed below as ambiguous.

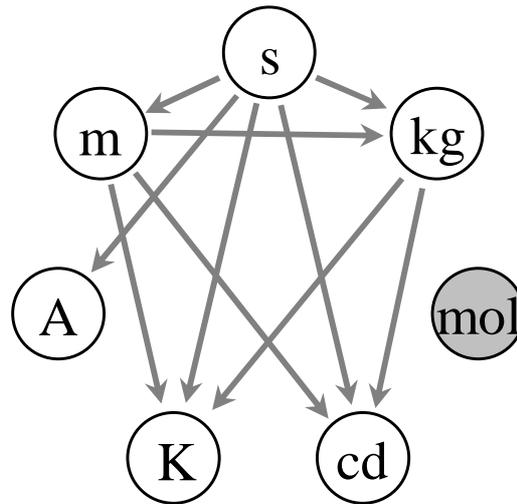

**Fig. 1.** Relations between the base units of the new SI system. The mole was adopted by majority vote at the Conférence Générale des Poids et Mesures (CGPM) in 1971 as the base unit of amount of substance. The mole is the only one SI base unit that was not the base unit in the original SI system in 1960. That is, it was already not necessary to be a base unit of the full system for some time.

## 3. Ratios

The integer numbers were analyzed in the previous section. A real number can be arbitrarily approximated as a ratio of integer numbers and an integer number is always a special case of real number. So, the real numbers that also express ratios will be treated in this section analogically, and thus the recommendation must also be the same.

The circumference $c$ can be expressed as

$$c = 2\pi r, \qquad (9)$$

where $\pi$ is Archimedes' constant, i.e. a purely mathematical constant and eq. (9) also corresponds to eq. (5). Note that $\pi$ can be evaluated experimentally e.g. using Buffon's needle and dimensionless probabilities. Hence, a measurement is not a sufficient condition for physicality. The circumference is a physical quantity of the same kind as the radius. The whole circle can be divided (graduated) into 360 angular degrees denoted as °. Then we can write

$$° \equiv \frac{2\pi}{360}. \qquad (10)$$

The minute of arc and the second of arc can be defined in a similar way as "sexagesimal" submultiples. However, their symbols represent only dimensionless numerical factors (fractions) contrary to their common use as units of the planar angle. Note that for historical reasons the symbols °, ′, ″ – corresponding to 0, 1, 2 – represented separators of the powers of "sexagesimal" submultiples used in writing a number that stood for a fraction of arc length (of the degree of arc or a different part of whole circle) or of time (hour or day) as well (i.e. as a fraction of different quantities). That is, it is an old mathematical notation of numbers and it is not a physical unit. For example, writing a numerical value of angle as 123°59′59″ without units does not mean that it is a product of angles 123° · 59′ · 59″ and their units. Note that the "sexagesimal" numeral system never used 60 symbols (numerical digits), and thus the separation symbols were introduced and misinterpreted later as a unit. A decimal separator or a thousands separator are not units as well. However, the symbol in eq. (10) is the mathematical factor that can be introduced as a factor and can be conventionally restricted to use for the planar angle where the decimalisation was not historically successful. However, such a restriction is also non-systematic as in the case of the Avogadro constant and it does not create a physical unit from a mathematical number because it is (should be) already included in the definition of the variable. The planar angle did not have units in the past (the radian and the steradian were introduced in the late nineteenth century). Therefore, it is possible to have the planar angle without units also in the future. The graduation of the curved line scale (e.g. on a protractor) as well as the graduation of the linear line scale (with an infinite curvature) represents the length quantity, while the planar angle means only a dimensionless fraction of a curved graduation length to a reference length. Angle measuring devices are historically based on arc divisions (curved coordinates). However, by dividing something into its parts does not change its dimension, and thus these sub-arcs are also length quantities. It is can also be seen from names of the inverse trigonometric functions, where the obtained quantity is "arcus" (the length of normalized arc) rather than the angle. Note that there is another non-systematic approach with a different degree. The degree Celsius, denoted as °C, was based on the centigrade scale between the melting point and the boiling point of water. The degree of Celsius (officially adopted in 1948) "is" same unit (with a different symbol) as the degree of Kelvin (1 °C = 1 °K). However, the temperature scale has a different offset (zero). Nevertheless, in other cases we do not use different symbols for units such as between different timescales. An offset definition must also be included in the definition of a quantity and it should not be transferred into units (e.g. "metres above sea level" are still the same metres; volts at a different ground potential are still the same volts).

  We can also find many examples of restrictions for the SI prefixes. The SI prefixes are multiples or submultiples of units, and thus they are also only numbers without a physical meaning. For example, "centi-" represents the real number 0.01 and its use is also restricted. It should not be combined with some units like "centisecond" or "centimillimetre". Nevertheless, there is no need for such restrictions from the perspective of the numerical result. The same numerical value 0.01 has another symbol, i.e. %. The percentage is dimensionless and it does not have the conventional restriction as the prefix centi- that it must be always used with a given physical unit and without another prefix. However, it does not make the percentage a unit as well as the angular degree a physical unit. Another example is the percentage point as a "unit" that is popular in finance and it is not explicitly rejected by the SI Brochure [7] to avoid the spread of such inconvenience. If we carefully define and describe the quantity, i.e. the ratio in this example, than the "point" is not needed and only the numerical factor equal to % can be used. Also non-decimal factors for calculations, e.g. the root mean square (RMS) or the peak-to-peak value, should not be included in units. For example, the RMS of a voltage (denoted as $U_{RMS}$) is correct (in volts). Nevertheless, the unit "volts RMS" (denoted as $V_{RMS}$) does not present a consistent approach. However, it is widespread in practice.

  The planar angle is formed by two rays (i.e. it is a mutual property between two half-lines) and supplemented with the corresponding sign. The sign for the measured planar angle above that includes an arc referring to the definition. However, it is only a figure that consists of lines (and of curved lines) measured in terms of a unit of length. That is, we cannot create an angle independently of lines or regions, if we define the value of the planar angle as an areal ratio of the circular sector to the disk like in a mathematical pie chart that illustrates numerical proportions generally (and not only angles). Note that the angle measuring device can also be calibrated by a second device using the ratio of the areas, if such devices are based on the Sagnac effect and share the wavelength of light and the angular rotation. However, a numeric ratio of two arbitrary lengths (areas) does not generally form an angle. The planar angle is a fraction that is restricted to e.g. a two lengths that must be specified. However, a restriction into a subset of fractions cannot create a physical unit from a number. The value of an angle is always calculated and it is not measured directly as a physical quantity.

  The problem with the dimension of the planar angle in physics is long-term [9]. The authors in [10] concluded that the idea that the angles are inherently length ratios is a misconception. However, it cannot imply, as they only concluded without a proof, that the planar angle is not inherently dimensionless. Their example with Brown's protractor showed that no other physical units are needed to be involved to define a measure of angles (i.e. mathematical fractions). That means, it is not necessary to have a length ratio, however, the planar angle can be defined as a length ratio (this was not disproved). It also suggests that angle is mathematical rather than physical, and thus it should not have a physical unit. The planar angle can be formally defined or calculated in many ways. The planar angle can be defined (derived) as the ratio of an arc length to a radius (mathematical approach) or the ratio of an arc length to another arc length (the circle closure techniques or the polygons that do not need a metrological traceability to the SI system when they are ideally realized) or the ratio of a length to another length using nonlinear cyclometric functions, such as e.g. arcsine (the sine bar) or arctangent (the tangent bar). And I can cite the VIM [2]: "The coherent derived unit for *every* derived quantity of dimension one in a given system of units is the number one, symbol 1. The name and symbol of the measurement unit one are generally not indicated." Additional reasons against such physical unit will be also presented in the next section.

The solid angle has a similar problem. Note that the special category was exclusively introduced for the radian and the steradian and it was named as the supplementary units of the SI system. However, the category was abrogated in 1995. It clearly shows a problematic nature of the existence of these units. Nevertheless, it is not a practical problem not to use such a unit, the steradian. The surface area of a spherical cap includes also the case of the surface area of the whole sphere. That is, they are quantities of the same kind with the same units. The radiant flux is a quantity that includes the power from all directions (the whole sphere). The radiant intensity is a quantity of power that comes through a specific spherical cap. That is, the detector physically measures the radiant flux (the power without a steradian) in both cases. The radiant flux of a source can be measured by an integrating sphere and the radiant intensity is a radiant flux restricted by a calculated dimensionless geometric factor (e.g. the surface area ratio of a spherical cap) represented by the part of the measurement device in front of the detector. This way, these quantities are different. However, they are quantities of the same kind with the same units (without the steradian) as well as the corresponding surface areas mentioned above. A similar example is an optical filter placed in front of the detector. The total flux (the power) and the transmitted power are two different quantities with the same units. The transmittance is a material factor that is dimensionless (expressed e.g. in % that is not a unit) as well as a dimensionless geometric factor mentioned above that is realized by an obstruction with a zero transmittance. By analogy, the radiosity and the radiance are quantities of the same kind. Note the similarity of the quantity names indicating the similarity in units. And by analogy, photometric quantities that often share the same symbols as their radiometric counterparts also do not need the steradian. The quantities must be distinguished (as well as the length of arc and the circumference). However, their units should not contain an extra unit for a dimensionless factor. For example, the transmittance $T = 0.3 = 30\ \% = 30 \cdot 0.01$. We do not need to introduce a unit like "relative" to avoid ambiguities by writing $T = 0.3$ rel. $= 30$ crel.

## 4. Identity

A special case of a number is one, denoted as 1. This number is generally equal to the identity element for the mathematical multiplication that is only one in common algebras.

In 1998, the Consultative Committee of Units (CCU) recommended an adoption of a new unit named "uno" representing 1 in dimensionless quantities that will be denoted as U. It can overcome the conventional restriction to use the prefixes only with some symbol of unit. However, it has not been adopted by the International Committee for Weights and Measures (CIPM) until now. The mathematical reason, why we should not introduce an additional symbol to the identity element, is ambiguity in writing. If

$$U = 1, \qquad (11)$$

then also

$$1 = U \qquad (12)$$

from the general symmetry of the equality relation. The ambiguity of "1" is known. It can be omitted as a factor or used in the arbitrary power $p$ as in the following identity

$$1 = 1^0 = 1^1 = 1^p, \qquad (13)$$

and thus the U also equals to its arbitrary power.

The special cases of the uno are the radian and the steradian. We can read in [7]: "However, it is a long-established practice in mathematics and across all areas of science to make use of rad = 1 and sr = 1. For historical reasons the radian and steradian are treated as derived units... It is especially important to have a clear description of any quantity with unit one...". That is, they are "unos" restricted to be used only for the planar and solid angles, respectively. Note that the higher-dimensional and the abstract spaces are also used in physics. For example, the Weinberg angle is used to describe a ratio of masses as well as the Avogadro constant in eq. (6). That is, the angle is generally too abstract (as well as a number) that it cannot be physical, and thus it cannot have physical units. If not, then there will be a need of infinite number of units for each dimensionality of space where hyperspheres exist. This seems to be impractical. And when only 2D and 3D angular units will be used, it is non-systematic. Moreover, the radian and the steradian are not commonly used in mathematics. And thus there is also no mathematical reason to use them in physics.

In this manner, the radian is also ambiguous in writing as it can be seen from the equation

$$\text{rad} = 1 = \text{rad}^p. \tag{14}$$

If the radian exists as the unit of angle that corresponds to the factor equal to 1 in eq. (9), then the angular frequency $\omega$ defined from the general frequency $f$ as

$$\omega \equiv 2\pi f, \tag{15}$$

will contain the radian. Then, the units of the angular momentum $L$ and the reduced Planck constant $\hbar$ will also contain the radian because of the classical definition

$$L \equiv I\omega \tag{16}$$

and

$$L = N_L \hbar, \tag{17}$$

respectively, where $I$ is the moment of inertia and $N_L$ is some numerical factor, and therefore they cannot contain the radian. It must be noted that the authors in [11] concluded that $I$ contains rad$^{-2}$ because they exclusively introduced their "angular radius of curvature", which contains rad$^{-1}$. However, the moment of inertia $I$ of any arbitrarily oriented object can be calculated by summations or by integrations in the Cartesian coordinate system because it corresponds to the second moment of the mass distribution, and thus no radian will appear in the result and likewise the radian will not appear in the multipole moments of the electric charge distributions or generally in any mathematical second moment corresponding e.g. to the variance in abstract spaces. For example, the moment of inertia for axis Z, denoted as $I_Z$, can be numerically calculated as

$$I_Z = \sum_i m_i \left(x_i^2 + y_i^2\right) \tag{18}$$

or

$$I_Z = \iiint \rho \left(x^2 + y^2\right) dx dy dz, \tag{19}$$

where the summation is over many point particles with masses $m_i$ and the integration uses the volumetric mass density $\rho$. We can clearly see that there is no place for the radian as it was derived for $I$ and unconventionally for many other quantities as it was suggested in [11].

Moreover, units of the angular kinetic energy $E_{ang}$ and the Planck constant $h$ will contain rad$^2$ due to eq. (15) because

$$E_{ang} \equiv \frac{1}{2} I \omega^2 \tag{20}$$

and

$$E_{ang} = hf, \tag{21}$$

respectively. It seems to be problematic with respect to the new SI definition of the kilogram, if the Planck constant contains the second power of the radian and it is considered as an unambiguous unit. On the other hand, we know that the classical linear kinetic energy is given by the equation

$$E_{lin} \equiv \frac{1}{2} m v^2 \tag{22}$$

and it does not contain the radian. Both energies are the quantities of the same kind, and thus we can add them to obtain the total energy. That is, we will sum quantities with a different power of the radian. This is a clear demonstration of the ambiguity of uno-like units that should not be used. The ambiguity of the radian was already mentioned e.g. more than 100 years ago in [12]: "It is the anomalous behaviour of the unit, radian ... This means we must insert or rub out the unit radian whenever it is convenient to do so." And I agree. Anomalies should be removed from the SI system.

Additional examples are trigonometric functions such as sine or tangent. They have argument $\varphi$ that represents the planar angle. In the calculations with such functions, the numerical value of an angle must be converted to the radian, and then the radian itself must be omitted in an argument, as it is equal to 1. If we use the Taylor series of the sine function that is often used as an approximation in physics (e.g. the sine bar or the sine error), we will obtain a polynomial of degree up to infinity

$$\sin(\varphi) = \varphi - \frac{\varphi^3}{6} + \frac{\varphi^5}{120} - \ldots \tag{23}$$

Analogically, the Taylor series of the cosine that are used e.g. for calculation of the period of a pendulum or the cosine error also lead to a higher-degree polynomial. However, the cosine error is approximated by $\varphi^2/2$. In the case of the radian as an unambiguous unit, the approximated expression of the cosine error has units equal to rad$^2$, which is not common for such dimensionless factor used for relative corrections. We see that a nonlinear function may give rise to problems with the use of a unit such as the radian. One cannot add together all different powers ("dimensions") of an "unambiguous" radian (as well as the angular degree as a unit). To avoid a necessity to use the formal metrological way of writing

$$\sin(\{\varphi\}) = \{\varphi\} - \frac{\{\varphi\}^3}{6} + \frac{\{\varphi\}^5}{120} - \ldots \tag{24}$$

and specifying that angles are in the radians, the unit of the planar angle that corresponds to the radian should be "equal" to one and the unit radian should be omitted. The solution is already common in mathematics. Thus, the second option in [10], i.e. dimensionless angles, seems to be correct and it can be easily applied. Imagine that something is rotating five-times per second. The frequency of full cycles $f$ is 5 Hz. The definition of quantity $f$ includes the cycle that should not be included (again) in the unit (e.g. abbreviated as "cyc"). The angular frequency $\omega \equiv 2\pi f = 10\pi$ Hz = $10\pi$ s$^{-1}$. The numerical value for $\omega$ is different only by the dimensionless multiplicative factor $2\pi$ that comes from the definition of angular frequency that is based on the quantity, eq. (15). However, the same unit should be used as in eq. (9). If not, the same problem will occur for mathematical functions, such as sin($\omega t$), that need their argument to be dimensionless.

Another argument against using units for the planar angle is the polar coordinate system. The polar coordinate system is a two-dimensional coordinate system. Note that a coordinate does not mean a dimension. The system uses two coordinates $r$ and $\varphi$ that can be converted to the Cartesian coordinates

$$x = r\cos(\varphi)$$
$$y = r\sin(\varphi). \tag{25}$$

An infinitesimal area element $dA$ can be calculated as

$$dA = dx \cdot dy = r \cdot dr \cdot d\varphi. \tag{26}$$

Thus, the planar angle $\varphi$ is a dimensionless coordinate (variable) to match the units from both sides of this equation. Even a clearer illustration provides the elliptic coordinate system with two dimensionless coordinates $\mu$ and $v$ where

$$x = a\cosh(\mu)\cos(v)$$
$$y = a\sinh(\mu)\sin(v). \tag{27}$$

with the focal distance equal to $2a$ and an infinitesimal area element equal to

$$dA = a^2 \frac{\cosh(2\mu) - \cos(2v)}{2} d\mu \cdot dv. \tag{28}$$

The unit equation corresponding to double-integrated elements from eq. (28) is

$$\text{m}^2 = \text{m}^2 \cdot 1 \cdot 1. \tag{29}$$

It obviously shows that the area with two dimensions of length is located in $a^2$ (multiplicative part of the Jacobian determinant) and the rest on the right-hand side of the equation must be dimensionless. We can also see that the elliptic coordinate system contains the hyperbolic functions as well as the circular functions. The arguments of the hyperbolic functions also do not have a unit different from 1. They are related to the circular functions through the imaginary unit $i$ that is purely mathematical without a physical meaning. A hyperbolic rotation corresponds to a circular rotation by an imaginary angle e.g. by

$$\cosh(\mu) = \cos(i\mu). \tag{30}$$

Thus, my suggestion not to use the radian as a unit also avoids a necessity of the imaginary radian.

The problem can also be demonstrated for a non-infinitesimal example where the infinitesimal area element can be replaced by a finite area element like a pixel (generally non-rectangular and position-dependent). Note that the pixel is an entity and it is not a measurement unit. The total area $A$ can be calculated by a summation of pixel areas. The summation can be done over relevant values of indices $i_x$ and $i_y$ (representing coordinates) as well as equivalently over a single index $i_{2D}$

$$A = \sum_{i_x} \sum_{i_y} A_{i_x i_y} = \sum_{i_{2D}} A_{i_{2D}}. \tag{31}$$

where all pixels are only renumbered by a different way. That is, the number of coordinates was changed without a change of dimension. Thus, the planar angle "units" as well as the pixels (and their indices) do not have units different from 1. The radian and the steradian should be omitted as well as 1 is also normally omitted because rad = 1 and sr = 1, this is "a long-established practice" as it is also mentioned in [7].

The problem is also related to the written form of equations. It was decided, for equations describing the measurement uncertainties to move from the numerical-value equations to the quantity equations. The quantity equation describes a relation with accompanying definitions of the quantities represented by their own symbols in the equation. This approach is also common in the scientific papers. However, the numerical-value equations are accompanied by statements specifying the units used in the equation. It is also the case of equations in [10], using statements e.g. "where $\theta$ is in degrees" and "where $\theta$ is in radians". However, the corresponding equations seem to be the quantity equations. Again, in order to avoid writing {$\theta$} "everywhere", it is better to treat the planar angle as a dimensionless factor without a physical unit.

## 5. Conclusions

The amount of substance and the angle should not have a unit different from 1. At least, they should not be a base unit of the SI system (if we accept them for some practical reasons). The integer numbers or real numbers should be kept numbers and they should not be defined as physical units. The approach to define the mole as a physical unit will wipe away the differences between the numerical values and the physical quantities. It is non-systematic to make units from non-unitary numbers and it is not formally correct. The historical reasons can be, e.g., a cause of the decimal number system. However, subjective conventions should be systematic as much as possible or practical to make the SI system easy, whereas the mole opens the door to define units for other restricted entities [13], like apples and oranges, because they cannot be derived within the SI system. Endless proposals such as "numerosity", "avo", "ent" are present [14], because they are based on their arbitrariness that has nothing with the physical reality. The removal of the mole from the SI units will stop it. Moreover, stoichiometric calculations in practice do not require the concept of the mole [14].

The radian has the ambiguity of uno-like units. The ambiguity can lead to an incorrect conclusion. For example, that the physical unit hertz is incoherent, while the radian is a coherent unit of the SI system of units [11] when we let escalate this ambiguity in formulae. The planar angle is generally defined as a ratio. That is, it is a dimensionless ratio because mathematics does not need physical units. The dimensionless quantities convey more information than a number. However, the information cannot be placed into the mathematical identity element creating an isolated base unit like that in Fig. 1. That is, we cannot incorporate the mathematical identity element into the SI system of physical units, and thus also the units of all relative ratios (e.g. ratios of prices, the Pearson correlation coefficients, probabilities, etc.).

The recommendation is not to use e.g. the angular degree and the mole as a physical unit. The value of an angle is a mathematical variable. The mole should be a numerical factor of a mathematical variable restricted to given entities and it should not be the base unit of the SI system. And for formal reasons it is also necessary to leave out the radian and the steradian in order to avoid the ambiguity that generally leads to mathematical uselessness. However, any future action will need deeper consensus emerging also from outside the metrological community to avoid problems related to such an important and broad issue. Nevertheless, the contradiction between the VIM and the new SI system definition must be resolved. We must balance between practical reasons (using e.g. μrad instead of $10^{-6}$) and the mathematical correctness. Nevertheless, it is relatively easy to remove all "rad" and "sr" from texts (the information is already present in the definitions of quantities) and such a removal will bring more practicality and also consistency within mathematics.

A general practical solution for writing could be the following. All quantities and units (already based as constant quantities [1]) with a physical dimension will be written using italic type. While numbers (numerals, the Avogadro constant or the mole), dimensionless variables (angles), dimensionless constants and SI multipliers will use roman type. It will also allow to use SI multiples without units that share some symbols (this sharing is a source of irregularities in the long term).